\documentclass[twocolumn,prl,footinbib]{revtex4-1}
\usepackage{amsmath}    % need for subequations
\usepackage{graphicx}   % need for figures
\usepackage{verbatim}   % useful for program listings
\usepackage{color}      % use if color is used in text
\usepackage{subfigure}  % use for side-by-side figures
\usepackage{hyperref}   % use for hypertext links, including those to external documents and URLs
\usepackage{float}
\usepackage{gensymb}
\usepackage{url}
\usepackage{xcolor}
\begin{document}

\title{Tuning of friction noise by accessing the rolling-sliding option}
\author{Soumen Das and Shankar Ghosh}
\affiliation {Department of Condensed Matter Physics and Materials Science, Tata Institute of Fundamental Research, Mumbai 400005, India}
\email{soumen.das@tifr.res.in , sghosh@tifr.res.in}

\begin{abstract}
		
	  Variable power transmission in mechanical systems is often achieved by devices, e.g., clutches and brakes, that use dry friction. In these systems, the variability in power transmission is brought about by engaging and disengaging the friction plates. Though commonly used, this method of making the coupling noisy is not as versatile as their electrical analog. An alternative method would be to intermittently vary the frictional force. In this paper, we demonstrate a self-organized way to tune the noise in the frictional coupling between two surfaces which are in relative motion with each other. This is achieved by exploiting the complexity that arises from the frictional interaction of the balls which are placed in a circular groove between the surfaces. The extent of floppiness in the coupling is related to the rate at which the balls make transitions between their rolling and sliding states. If the moving surface is soft and the static surface is hard we show that with increasing filling fraction of the balls the transitions between rolling and sliding against the static surface give way to the transitions between
rolling and sliding against the moving surface. As a consequence, the noise in the coupling is large for both small and large filling fraction with a dip in the middle. In contrast, the sliding with the static surface is suppressed if the moving pate is hard and the noise in the coupling decreases monotonically with the filling fraction of the balls.
		
\end{abstract}

\maketitle

Dry friction is commonly used as a coupling mechanism to transmit power in mechanical systems. Examples of this can be seen in automotive vehicles where the friction plates are used for clutching and braking purposes \cite{sclater2001mechanisms,orthwein2004clutches}.  If a sudden brake is to be applied, it is preferable that the coupling between the brake pads and the inner rim of the wheel is strong. Similarly, for maximum force transmission, the clutch should strongly couple the gearbox to the engine. However, when caught up in traffic while driving uphill, one often uses a technique called \textit{feathering the clutch} or \textit{slipping the clutch} \cite{DrivingTheEssentialSkills} where a driver gets better control over the vehicle by alternately pressing and releasing the brake or clutch which makes the frictional coupling time-varying (noisy).  In essence, there are situations that may demand a  mechanical system to exhibit strong frictional coupling in one instance of time and weak frictional coupling in another instance.   This is commonly achieved by making the coupling noisy where the system is constantly made to toggle between states which have strong (e.g. brakes on) and weak (e.g. brakes off) frictional coupling. In mechanical systems, the presence of inertial forces makes this general principle of controlling the power transmission by varying the duty cycle more difficult to implement as compared to the pulse width modulation technique (PWM) employed in their electrical analogs \cite{holmes2003pulse,sun2012pulse}.  In addition, it causes fretting and associated mechanical failure, e.g.,  continuous driving with a feather clutch technique will quickly destroy the clutch.  In what follows we will call the coupling noisy if the frictional force toggles constantly between strong and weak coupling states.  In the experiments reported here, we show ways in which the complexity that arises from the third body frictional interactions of balls sandwiched between two surfaces can be harnessed in order to tune the noise in the frictional coupling.

In our experiments  we place millimeter-sized  balls on a circular groove between a static and a moving plate. These balls, which are constrained to move in single file, mediate the frictional drag exerted by one plate on the other. The coupling between the top and the bottom plate which strongly depends on the dynamics of the balls is measured in terms of the spread in the coefficient of friction. At a small filling fraction each ball exhibits periodic rolling and sliding against the hard plate. The toggling between rolling and sliding generates a spread in the coefficient of friction. With increasing filling fraction, the sliding mode gets progressively suppressed. This suppresses the noise in the coupling between the plates. For a soft moving plate, this decrease in noise is non-monotonic. At higher filling fraction, the collective dynamics of forming and breaking of clusters set in due to the sliding against the soft plate. This restores the noise in friction coefficient. Such non-monotonic variation in noise is absent if the moving plate is hard and the static plate is soft.

\begin{figure}[t]
	\centering
	\includegraphics[width=1\linewidth]{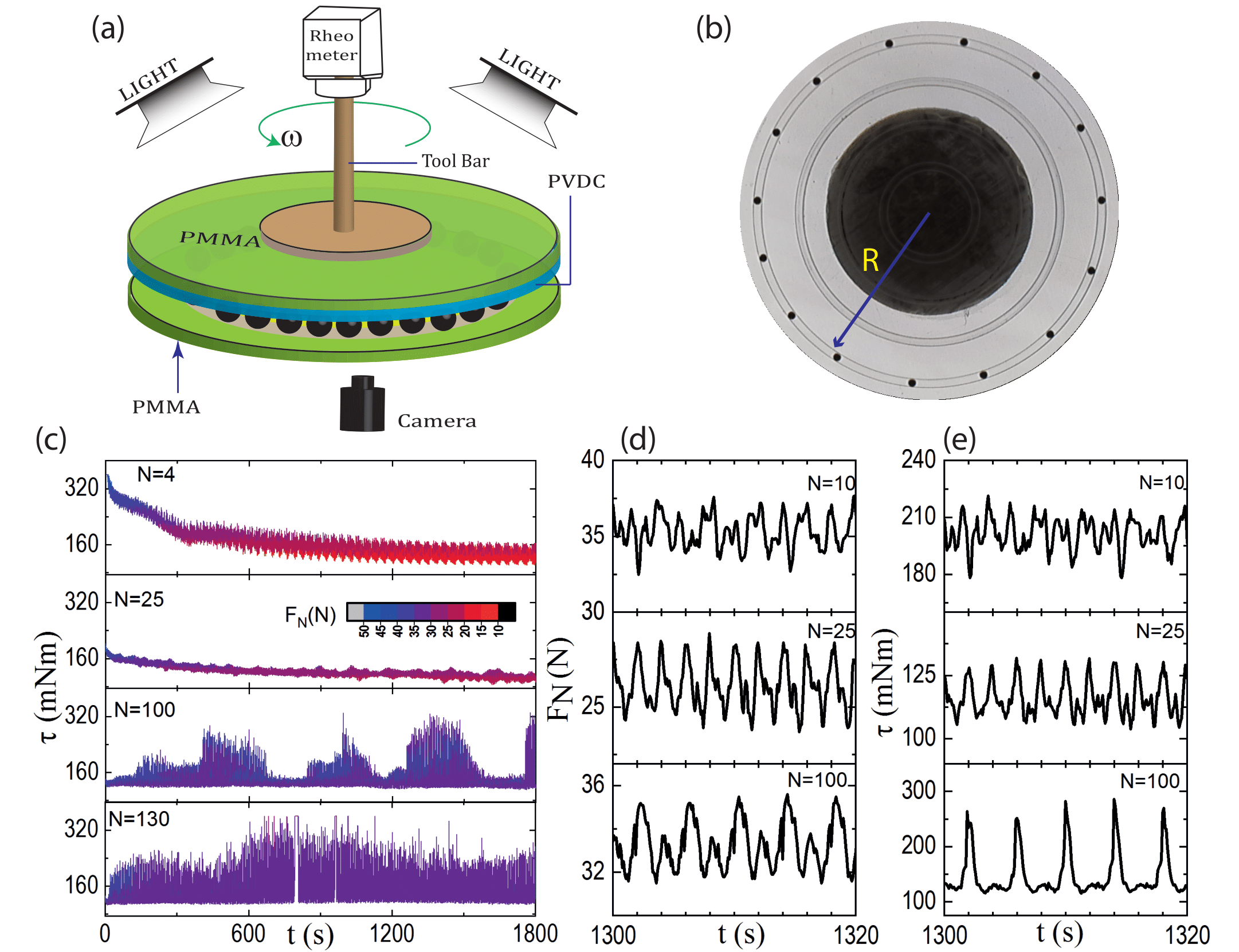}
	\caption{(a) Schematic of the experimental setup in soft-top and hard-bottom plate geometry. A parallel plate toolbar of the rheometer is attached to the top PMMA disk and the soft PVDC circularly cut sheet is attached to it. The radius of the track is $R=47$ $\rm mm$. Initially, the balls are uniformly distributed on the track and the top plate is made to exert a normal force $F_N=30$ $\rm N$ on them. The balls are set in motion by rotating the top plate at a uniform angular velocity $ \omega = 180$ $\rm \deg/s$.   The camera is placed below the bottom PMMA plate while the illumination is from the top. The image in (b) captured by the camera shows the initial configuration of the balls on the groove. The central darker portion in the region of attachment of the measuring toolbar with the PMMA disk. (c) Time trace of torque $ \tau $ for some representative values of $N$. Normal force $F_N$ is color-coded. (d) and (e) show the zoomed-in part of the time series of $F_N$ and $ \tau $.
	}
	\label{fig:fig1}
\end{figure}

The experimental setup is shown in Fig. \ref{fig:fig1} (a).  A circular plate of transparent Polymethyl methacrylate (PMMA) of $60$ $\rm mm$ radius and $6$ $\rm mm$ thickness is used as the bottom plate into which an annular groove of inner radius $R= 47$ $\rm mm$ and width $ 1.9$ $\rm mm $ is carved. The groove has a depth of $0.6$ $\rm mm$ and it acts as a circular track for the balls to move. $N$ stainless steel balls of diameter $2$ $\rm mm$ are placed on the groove.  They are set in motion by rotating the top plate which is coupled to a rheometer (Physica MCR-301) via a shaft. The experiment is performed in two geometries. In the first geometry, the top plate is constructed by attaching a circular transparent Polyvinylidene chloride (PVDC) polymer sheet (see SI \cite{supplementary}) of thickness $1$ $\rm mm$ and radius $55$  $\rm mm$ to the bottom surface of a similar-sized circular plate made from PMMA (soft-top and hard-bottom plate geometry). In the second geometry, the top plate is made from PMMA while the  bottom  plate  is made by covering the grooved PMMA plate with a  PVDC sheet (hard-top and soft-bottom plate geometry). When the  plates are pressed against the balls, the soft PVDC sheet deforms around each ball.  The tensile stress developed in the polymer sheet pushes the balls away from each other \cite{johnson_1985} (see SI \cite{supplementary}) which reduces the inter-ball friction and allows us to run the experiments over a long period of time.     

We first describe the experiment in soft-top and hard-bottom plate configuration. Initially, the balls are uniformly distributed on the track and the top plate is made to exert a normal force $F_N=30$ $\rm N$ on them. Then the top plate is set to rotate at a uniform angular velocity $ \omega = 180$ $\rm \deg/s$  while maintaining a constant average gap with the bottom plate. A rheometer is used to monitor both the normal force (resolution $2$ $\rm mN$) and the torque $\tau$ (resolution $1$ $\rm nNm$) acting on the top plate to maintain its set angular velocity at every 100 ms. Additionally, configurations of the balls on the track are continuously imaged at a rate of $3$ frames per second. We perform the experiments by varying the filling fraction of balls, $\rho_{N} = \frac{N}{N_T}$,  where $N_T = 150$ is the total number of balls that can be fit on the groove.

The balls can access two motional states - (i) rolling and(ii) sliding \cite{kumar2015granular,ghosh2017geometric,ghosh2018geometric}. They will be referred to as `rollers' and `sliders' respectively. For the `rollers', the top plate moves with a relative velocity $\frac{R\omega}{2}$ with respect to the balls. The `sliders' themselves can be of two types: (i) `B-sliders' - which slide with a velocity $R\omega$ with respect to the bottom plate and are at rest with respect to the top plate and (ii) `T-sliders' - which slide with respect to the top plate and are at rest with respect to the bottom plate. The various spatio-temporal configuration of these `rollers' and `sliders' constitute the internal frictional states of the system and transitions between them generate noise in the friction coefficient. Each of these states is characterized by different frictional forces. We will identify these states from the analysis of the spatio-temporal configurations of the balls later. The force of friction associated with the `rollers' is lower than that associated with the `sliders'. In the context of soft-top and hard-bottom plate geometry, the force of friction associated with `T-sliders' is significantly larger than that associated with `B-sliders' (see SI \cite{supplementary}).

\begin{figure}[t]
	\centering
	\includegraphics[width=1\linewidth]{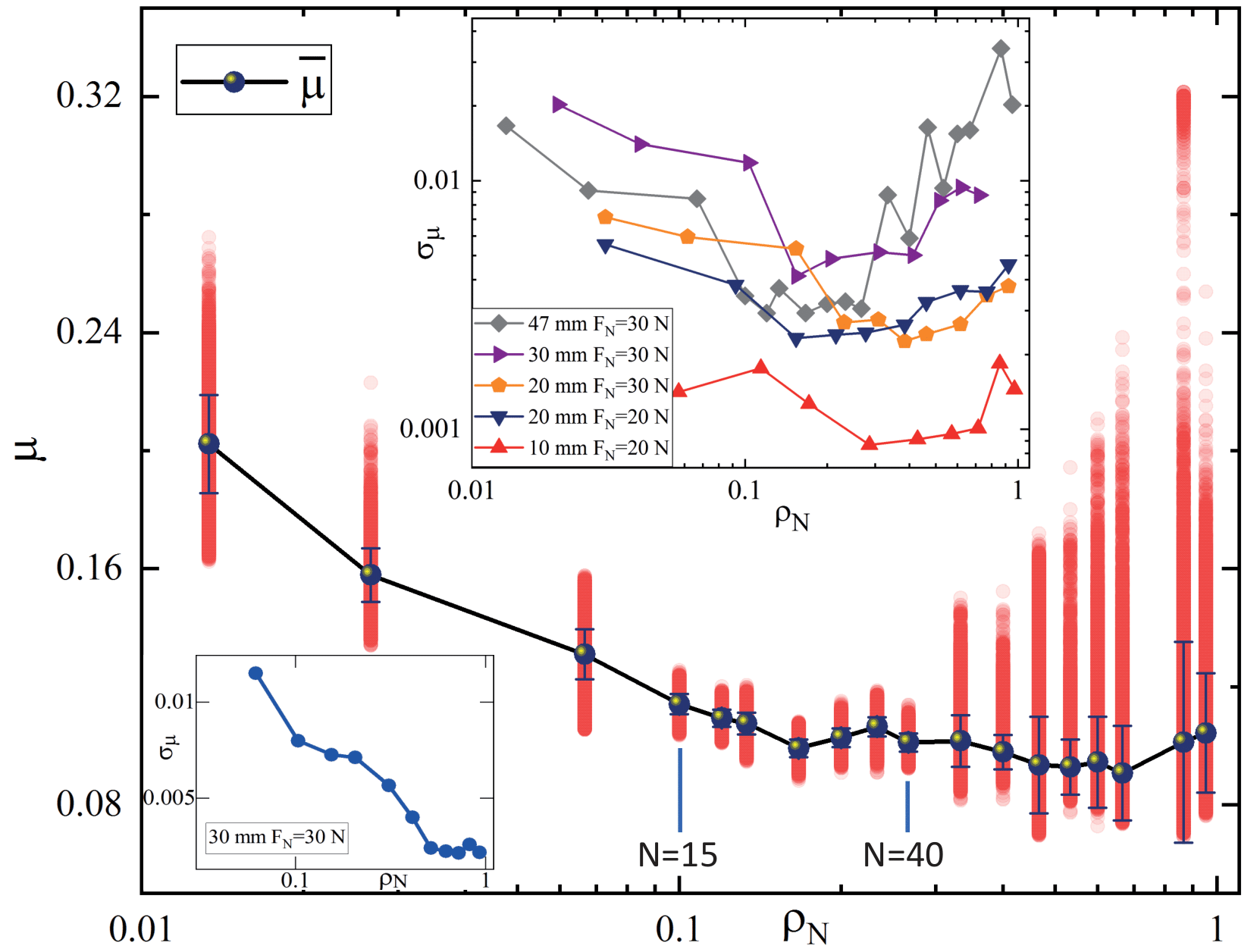}
	\caption{The scatter plots correspond to $\mu$ as a function of $\rho_N$. The extent of spread in $\mu$ is particularly low between $\rho_N=0.1$ $(N=15)$ to $\rho_N=0.266$ $(N=40)$. Black dots show $\bar{\mu }$ decreases with $\rho_N$. Top right inset shows $\sigma_{\mu}$ as a function of $\rho_N$ for different track sizes ($R$) at different initially applied normal forces in soft-top and hard-bottom plate geometry. The bottom left inset shows the same plot in the case of hard-top and soft-bottom plate geometry.
	}
	\label{fig:fig2}
\end{figure}	

The normal force between the top plate and the balls depends on the extent of deformation made by the balls into the top plate. Due to unintended machining errors and lack of parallelism between the top and bottom plates, the gap height between them has an angular profile (see SI \cite{supplementary}). Crests and troughs in the gap profile are not always occupied. As the balls move into/away from these regions, normal force exhibits variation in time which then triggers toggling between the internal frictional states and produces a change in torque [Fig. \ref{fig:fig1} (d) and (e)]. Figure \ref{fig:fig2} shows the variation of the coefficient of friction, $\mu=\frac{F_f}{F_N}=\frac{\tau}{R.F_N}$ as a function of $\rho_N$, where $F_f$ is the friction force acting on the top plate. Clearly, the extent of spread in $\mu$ changes non-monotonically with $\rho_N$. The spread in $ \mu $ is a measure of the noisiness in the coupling between the plates. Larger the spread, noisier is the coupling. This is particularly evident for small and large $ \rho_N $. However, for an intermediate-range, $ 0.1 \le \rho_N \le 0.266 $, the extent of variation in $ \mu $ is strongly suppressed. 

\begin{figure}[t]
	\centering
	\includegraphics[width=1\linewidth]{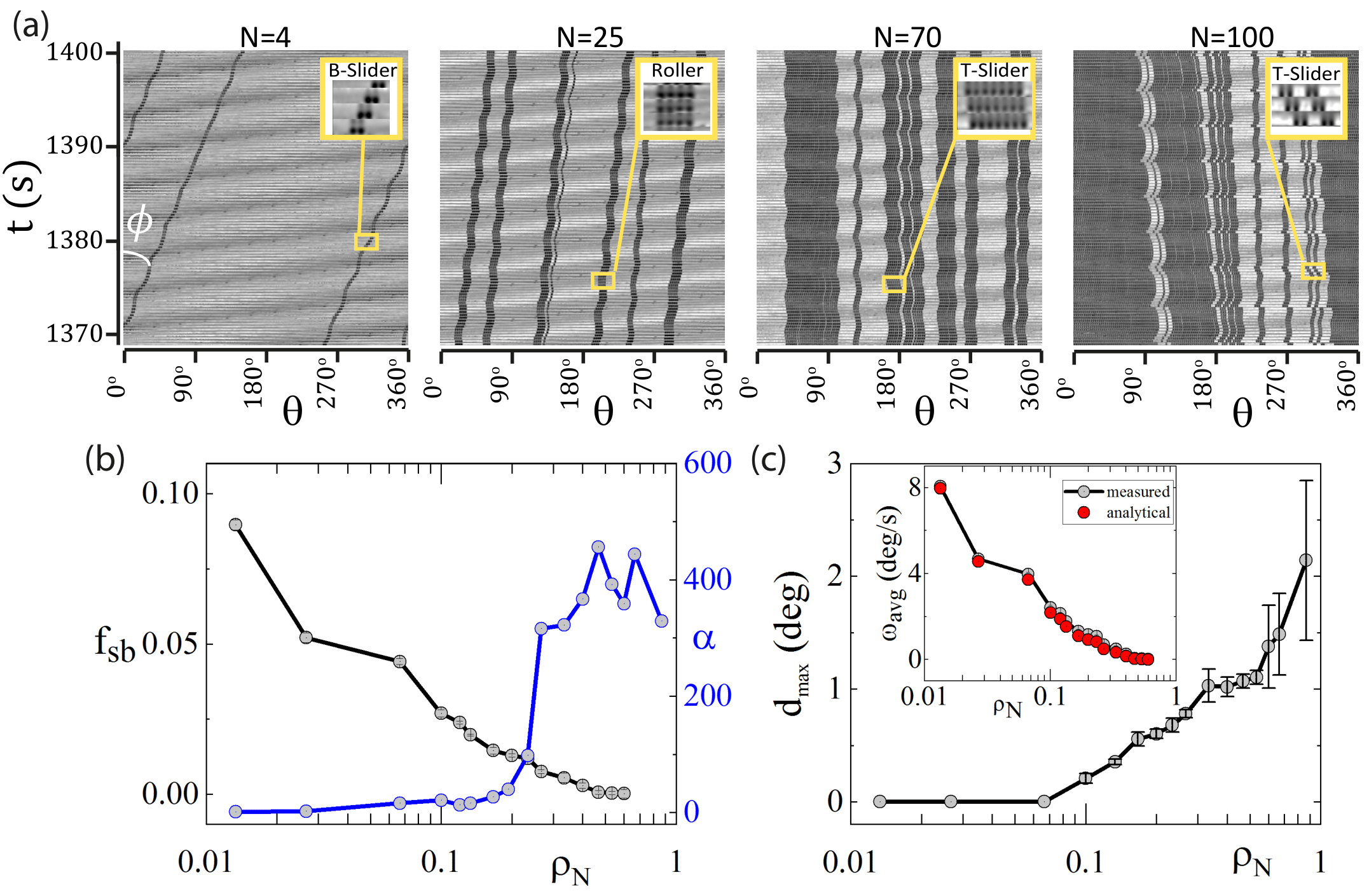}
	\caption{The circular groove is mapped onto a straight line. Pure rolling component of motion is subtracted from the images and they are stacked on top of each other. Different states of the balls are shown in the inset. The left $y$ axis of the dual plot in (b) shows the monotonic decrease of $f_{sb}$ with $\rho_N$ (shown in black). Points beyond $\rho_N \geq 0.666$ are not shown as sliding is stopped. $\alpha=\sum_t \left| Z_{\rho_N}(t)-Z_{\rho_N}(t+\delta t)\right|$ is a measure of incoherent dynamics which is shown in blue [right $y$ axis in (b)]. Here $\delta t=1/3=0.33$ s. (c) Shows the largest slippage events $d_{max}$ averaged over a time interval of $0.33$ s. Inset shows the variation of the average velocity $\omega$  with the filling fraction $\rho_{N}$ in the `R-frame' of reference. }
	\label{fig:fig3}
\end{figure}

The black dots in Fig. \ref{fig:fig2} correspond to the measure of the mean value of the friction coefficient $\bar{ \mu }$ which decreases with increasing $\rho_N$. This drop is related to the concomitant decrease in the fraction of `B-sliders', a feature of the experiment that we will discuss next. We have performed the experiments for different sizes (radius $R$) of the track at different initially applied normal forces and all of these show a dip in the standard deviation in the coefficient of friction, $\sigma_{\mu}$ as a function of $\rho_N$ (top right inset of Fig. \ref{fig:fig2}). 		 

To establish the correlation between the configurations of the balls on the groove and the measurements of friction coefficients, the circular groove is mapped onto a straight line. To figure out the mode of motion of the balls, the pure rolling component of motion is subtracted from the images. These transformed images are referred to be in the `R-frame' of reference. These images are then stacked on top of each other to create a montage [Fig. \ref{fig:fig3} (a)]. If $\phi$ is the angle of a trajectory with the vertical, then the average velocity can be obtained as $\omega_{avg}=\langle \tan\phi \rangle$. In the `R-frame', `rollers' move parallel to the vertical.  For `B-sliders', the trajectory is at an angle $tan^{-1}\left(\frac{\omega}{2}\right)$ with respect to the vertical towards right since the velocity is twice that of pure rolling. While `T-sliders' are at an angle $-tan^{-1}\left(\frac{\omega}{2}\right)$ with respect to the vertical. Therefore, $\omega_{avg}=f_{sb}.\left(\frac{\omega}{2}\right)+f_r.0-f_{st}.\left(\frac{\omega}{2}\right)$, where $f_{sb}$, $f_r$ and $f_{st}$ are the fractions of time the balls spend as `B-sliders', `rollers' and `T-sliders' respectively and $f_{sb}+f_r+f_{st}=1$. It can be observed from Fig. \ref{fig:fig3} (a) that for small $\rho_N$, fraction of `B-sliders' is more. This happens due to large normal force per ball and large frictional grip of the soft top plate on the balls. As $\rho_N$ is increased, normal force per ball decreases and balls become more of `rollers'. However, with decreasing normal force per ball the extent of deformation in the soft plate reduces which in turn weakens the repulsive interaction between the balls. Thus occasionally balls can touch each other and get jammed momentarily. Hence top plate slip past them and generate `T-sliders'. For large $\rho_N$ ($N\geq100$ or $\rho_N\geq 0.666$) the system has an overall small sliding component with respect to top plate ($|\tan\phi| \leq 0.1$ $\rm \deg /s$). We have neglected $f_{st}$ while considering the motion of the balls for $\rho_N < 0.666$. Hence $f_{sb}=\omega_{avg}/\left(\frac{\omega}{2}\right)$ and $f_r=1-f_{sb}$. Left $y$ axis in Fig. \ref{fig:fig3} (b) shows the monotonic decrease of $f_{sb}$ with increasing $\rho_N$. Assuming the amplitude of the individual torque signals due to `B-sliders' and `rollers' as $\tau_{sb}$ and $\tau_r$ respectively, the rms value of the net torque signal arising from the periodic toggling between `B-sliders' and `rollers' is $\tau_{rms} = \sqrt{\tau_{sb}^2 D+\tau_{r}^ 2(1-D)}$,where $D=f_{sb} /f_{r} \approx  f_{sb}/ (1- f_{sb})$ is the duty cycle.  Since $\tau_{rms}$ is a monotonically increasing function of $f_{sb}$, the noise in the coupling reduces with decrease in $f_{sb}$ for $\rho_N <0.666$.

In the `R-frame' of reference, `B-sliders' contribute to the particle current which decreases with decrease in $f_{sb}$ [grey symbols in inset of Fig. \ref{fig:fig3} (c)]. The average velocity of the balls in this frame can be calculated analytically by modeling the particle transport in terms of a totally asymmetric simple exclusion process (TASEP) with periodic boundary conditions \cite{SCHUTZ20011,Kriecherbauer_2010}.  This is a reasonably good model if we do not consider the occasional `T-sliders' which are very small even in case of higher $\rho_N$. Here we have assumed that for a given $\rho_N$, all balls have the same hop rate $p$ which is determined by $f_{sb}$. The average velocity is then given using random sequential update rule as $\omega_{avg}=\left(\frac{\omega}{2}\right). \frac{J}{\rho_N}=\left(\frac{\omega}{2}\right). p(1-\rho_N)$ \cite{Rajewsky1998,Kanai_2006,Kanai_2007}, where $J$ is the average current in the system, $ p=f_{sb}$ and $\frac{\omega}{2}$ is the normalization constant. Inset of Fig. \ref{fig:fig3} (c) shows that measured and analytical average velocities are in good agreement with each other.

\begin{figure*}[t]
	\centering
	\includegraphics[width=.75\linewidth]{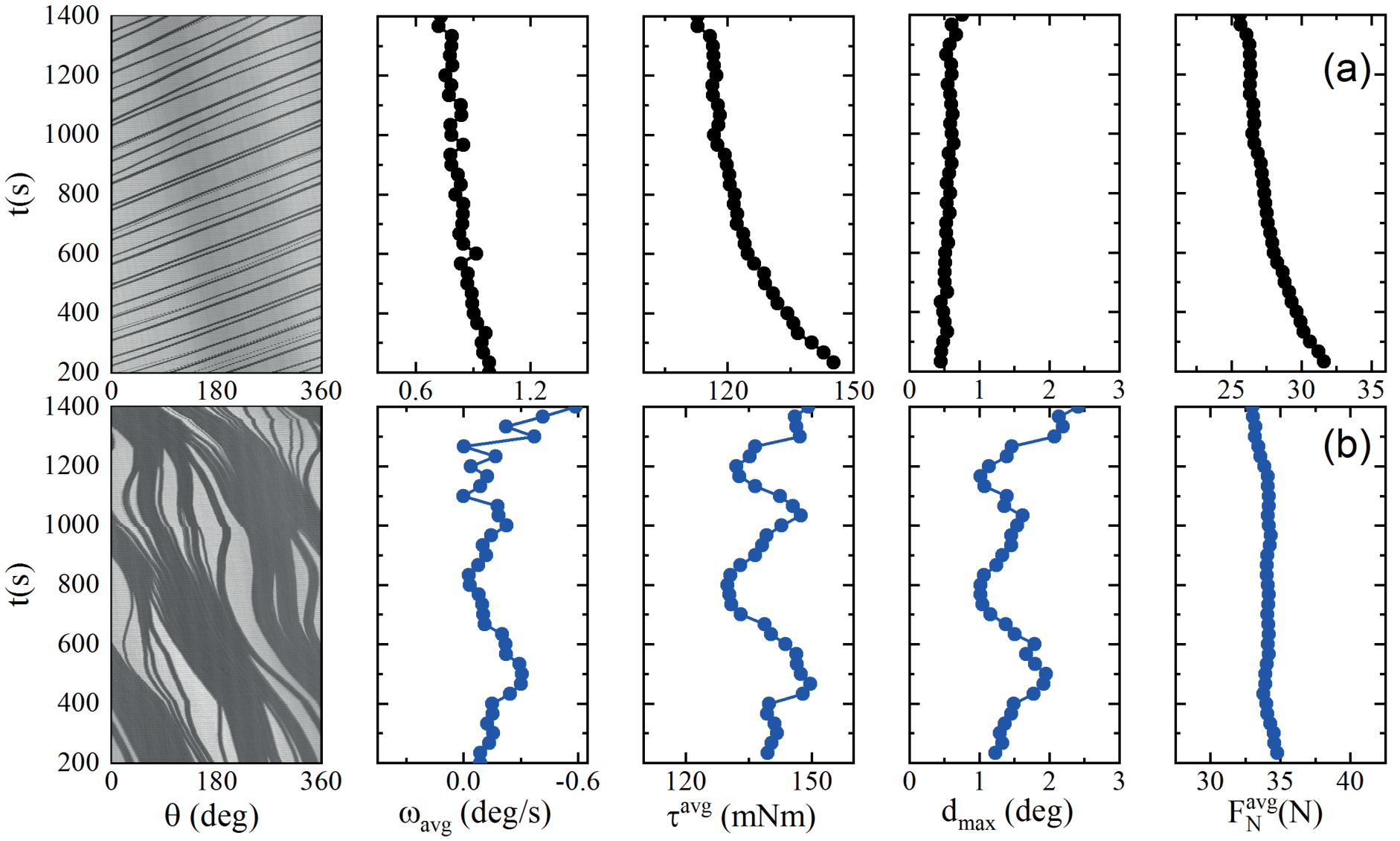}
	\caption{Trajectories and averaged transport data for (a) $N=25$ and (b) $N=100$ balls. The averages are calculated over a time interval of $0.33$ s. }
	\label{fig:fig4}
\end{figure*}

In addition to the monotonic decrease in  $f_{sb}$ and $\omega_{avg}$  with $\rho_N$, the system also shows a transition from a coherent to an incoherent dynamics as a function of $\rho_N$. The fluctuation in the size of the clusters is one measure of such dynamics which is calculated in terms of the quantity $ \alpha=\sum_t \left| Z_{\rho_N}(t)-Z_{\rho_N}(t+\delta t)\right|$, where $Z_{\rho_N}(t)$ is the total number of clusters in the system at time $t$ and $\delta t=1/3$ s. The data is shown in blue in the right $y$ axis in Fig. \ref{fig:fig3} (b). For small $\rho_N(\leq 0.2)$, the system exhibits coherent dynamics where the clusters do not evolve and all of them keep moving with a nearly constant average velocity [e.g. Fig. \ref{fig:fig4} (a)]. Whereas for larger $\rho_N$, the clusters constantly exchange particles. This results in a considerable variation in average velocity, an example of which is shown in Fig. \ref{fig:fig4} (b). Note that the trajectories have a negative slope [left panel of Fig. \ref{fig:fig4} (b)], which corresponds to the top plate slipping with respect to the balls. This happens when there is a transient jam which results in a sudden drop in the mobility of the balls. The slippage events assist the breaking of the clusters by generating elastic disturbances in the top plate. The largest slippage event in a given time interval ($d_{max}$) is strongly correlated with the incoherent dynamics of the cluster coalescence and fragmentation [Fig. \ref{fig:fig3} (b) and (c)] and the averaged torque ($\tau^{avg}$) in that time interval [third and fourth panel inFig. \ref{fig:fig4} (b)]. It should be noted from Fig. \ref{fig:fig4} (b) that the normal force remains mostly silent, i.e., the slowly varying changes in the friction coefficient comes mainly from variations in the torque.

The two above mentioned contributions, i.e., decrease in the sliding of the balls with respect to the bottom plate and increase in the incoherent dynamics of the clusters at large $\rho_N$ [Fig. \ref{fig:fig3} (b) and Fig. \ref{fig:fig4} (b)] compete with each other to make the noise in $\mu$ to have a non-monotonic dependence on $\rho_N$.

In the case of hard-top and soft-bottom plate geometry, the bottom plate exerts a larger frictional grip on the balls and hence `B-sliders' are always absent. In this scenario, the noise in the coupling arises due to the toggling between `T-sliders' and `rollers'. With increasing $\rho_N$, $f_{st}$ and hence $\sigma_{\mu}$ decreases monotonically with $\rho_N$ (bottom left inset of Fig. \ref{fig:fig2}). This is due to the following observation. When $ \rho_N $ is small, the balls are mainly `T-sliders' as they are held in their place by the large  
\begin{figure}[h]
	\centering
	\includegraphics[width=1\linewidth]{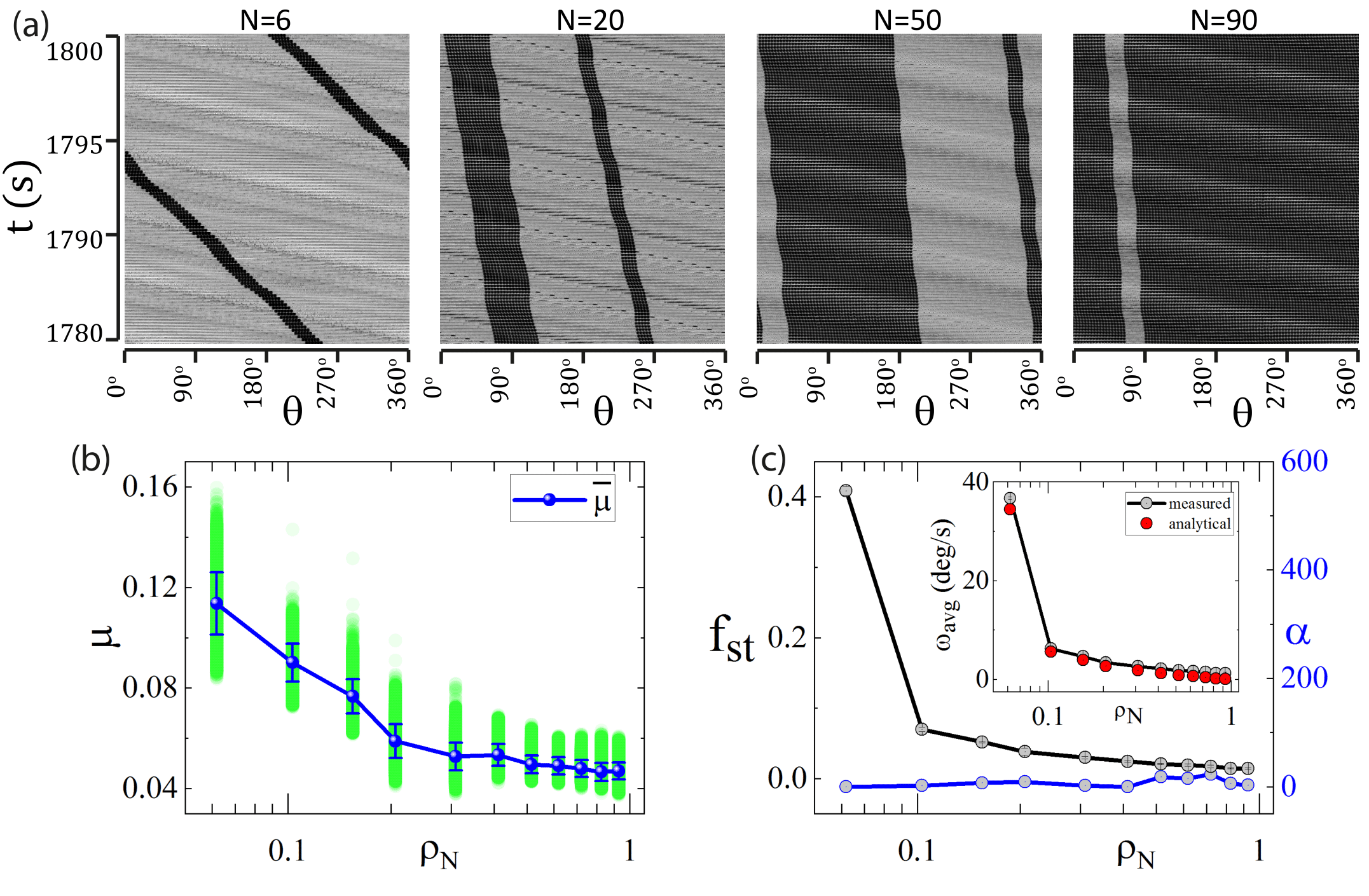}
	\caption{(a) Montage of images in the case of hard-top and soft-bottom plate geometry. 
			 (b) Spread in $\mu$ decreases monotonically in this case. The left y-axis of the dual plot in
			 (c) shows the fraction of sliding with respect to the top plate as a function of $\rho_N$ (shown in black). $\alpha$ is shown in blue in the right y-axis. Inset in (c) compares the measured and analytical average velocities.  }
	\label{fig:images_rev}
\end{figure}
  deformation generated in the soft bottom plate. With increase in $ \rho_N $, the deformations reduce and the balls begin to roll more (Fig. \ref{fig:images_rev}). In the absence of `B-sliders' which slip with respect to the soft plate to generate the elastic disturbances that cause fragmentation of the clusters, the dynamics become coherent for all $\rho_N$.  The fractions $f_{st}$ and $f_r$ can be determined as $f_{st}=\omega_{avg}/\left(\frac{\omega}{2}\right)$ and $f_r=1-f_{st}$. In this case, $f_{st}$ decreases monotonically with $\rho_N$ (Fig. \ref{fig:images_rev} (c) left y-axis). Toggling between `T-sliders' and `rollers' contributes to the noise in the coupling. The noisiness in the coupling reduces with a decrease in $f_{st}$. Whereas $\alpha$ which is the measure of incoherent dynamics does not change at all [Fig. \ref{fig:images_rev} (c) right y-axis] and all the clusters move with a constant velocity. Hence there is no contribution coming from the exchange dynamics of the clusters to compete against the reduction of noise in $\mu$ due to a decrease in $f_{st}$ with $\rho_N$.

In addition, we utilize the recently introduced idea of using lossless data compression \cite{PhysRevX.9.011031} to verify the structural correlation in this system. The detailed analysis and results are provided in the supplementary section \cite{supplementary}.

Frictionally coupled objects when driven tend to get jammed \cite{liu1998nonlinear,bi2011jamming}.  
To unjam the system, it is necessary to periodically inject energy into it. This makes the resulting motion intermittent. In this paper, we demonstrated a route to tune the extent of this intermittency in the third-body friction \cite{godet1984third,godet1990third,singer1998third,deng2019simple} and via it gain control over the noise in the frictional coupling between two surfaces. In the experiments reported here, steel balls which form the third-body that is sandwiched between a static bottom plate and a rotating top plate, are made to move in single file on a circular track. The noise in the coupling between the plates is measured in terms of the spread in the friction coefficient. Our experimental system is inspired by the mechanical arrangement of a ball bearing. In a conventional ball bearing, balls made from hard materials are sandwiched between the bearing's inner and outer races. Each ball is held in its place by means of a cage. This ensures that during motion the balls do not collide with each other and get frictionally jammed. Though the collective dynamics of the balls are suppressed in a ball bearing, it can still exhibit noisy dynamics because of the presence of play in its mechanical couplings \cite{mevel1993routes, KAHRAMAN1991469}. In contrast, we do not hold the balls in cages and the noise observed in running our experimental system is related to the collective dynamics of these balls. As a closing remark, we would like to point out that the conventional wisdom that the coefficient of friction is approximately constant for a pair of surfaces \cite{ghosh2017geometric} can easily be circumvented by harnessing the complexity of the third body interaction forces. 

We acknowledge support of the Department of Atomic Energy, Government of India, under Project No. 12-R\& DTFR-5.10-0100.

\end{document}

\title{Supplementary Material for \\``Tuning of friction noise   by accessing the rolling- sliding option''}
\author{Soumen Das and Shankar Ghosh}
\affiliation {Department of Condensed Matter Physics and Materials Science, Tata Institute of Fundamental Research, Mumbai 400005, India}
\email{soumen.das@tifr.res.in , sghosh@tifr.res.in}

\maketitle

\begin{figure}[h]
	\centering
	\includegraphics[width=.75\linewidth]{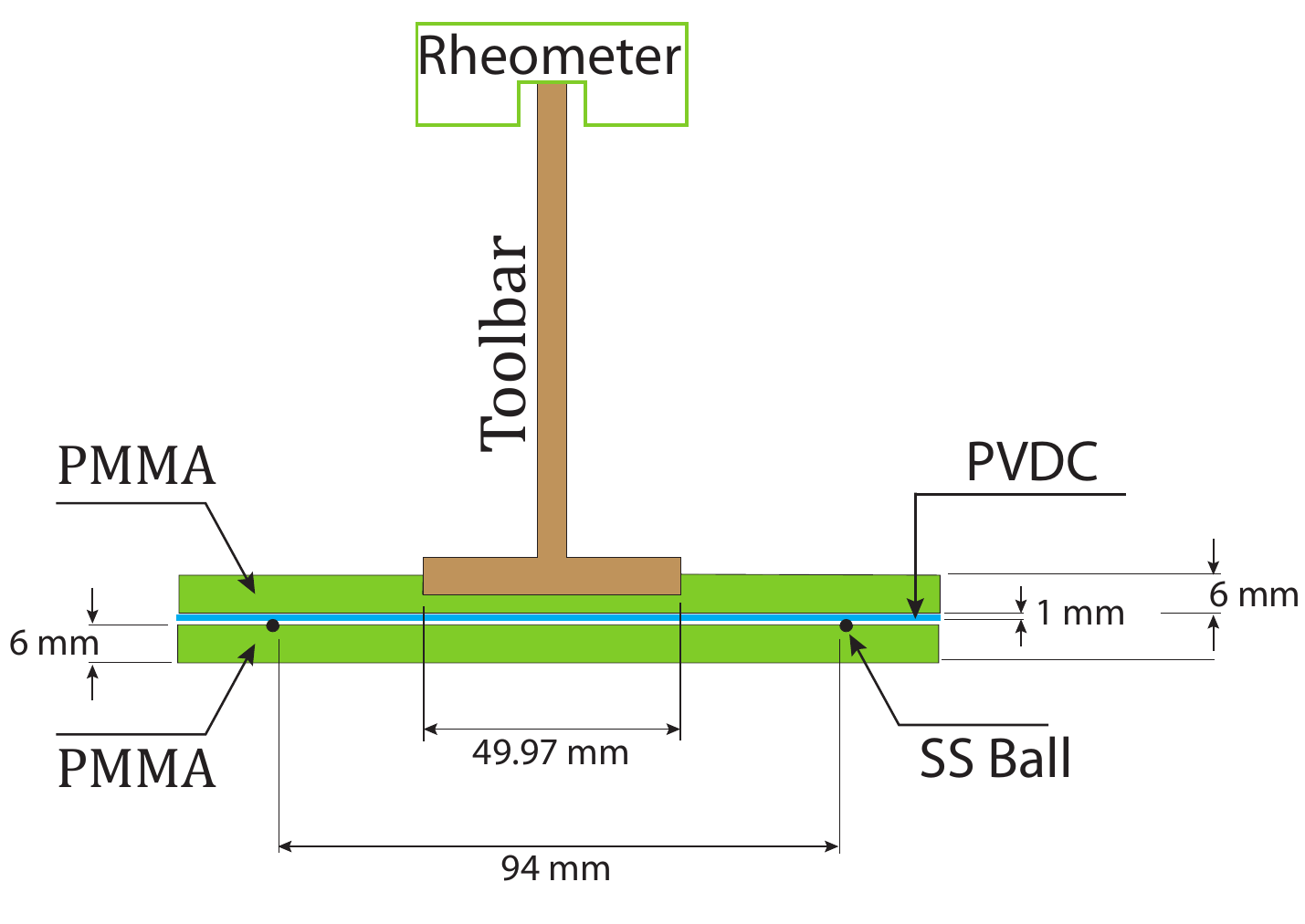}
	\caption{\textbf{Experimental Details}: The figure shows the cross-section side view of the experiment in soft top-hard bottom plate geometry. We use the commercial rheometer Physica MCR-301.  We use grade 1.4401 stainless steel balls of diameter $2$ $\rm mm$. As a bottom plate, we use transparent Polymethyl methacrylate (PMMA) disc of thickness $6$ $\rm mm$ and radius $60$ $\rm mm$ cut using a laser cutter. An annular groove of inner radius $ 45$ $\rm mm$ and width $ 1.9$ $\rm mm $ is carved on it. The groove has a depth of $0.6$ $\rm mm$. PMMA has Young's modulus of about $3$ $\rm GPa$.	The reason for using one softer plate is as follows. If both the plates are hard, then clustering can cause the inter-ball contact friction to increase to a value such that the torque supplied by the rheometer is no longer sufficient to keep the top plate rotating at a constant angular speed. In such a case the system begins to generate screeching sounds and finally comes to a halt. The contact force between the balls can be lowered by using a softer plate. The elastic deformation of the soft palate creates repulsion between neighboring balls. This is explained in detail later.	The soft and transparent Polyvinylidene chloride (PVDC) sheets that we use are industrially made by gluing three  $ 333$ $\rm  \mu m $ thick PVDC sheets. These sheets are usually used as curtains in industrial settings.  The glue can be removed by immersing the composite thick PVDC sheet in chloroform. Chloroform being a poor solvent for PVDC does not affect the individual sheets. 	The individual $ 330$ $\rm \mu m $ thick sheets tend to tear during the experiments while the thicker sheets (made by gluing multiple thin sheets) were found to have large variations in thickness. Thus, as a trade-off we use $ 1$ $\rm mm $ thick commercially obtained PVDC sheets.	The soft PVDC sheet of radius $55$ $\rm mm$ is attached to the bottom surface of a circular plate made from PMMA of thickness $6$ $\rm mm$ and radius $55$ $\rm mm$. Young's modulus of the PVDC sheet is $\sim 15$ $\rm MPa$. The PVDC sheet is attached to the PMMA plate by holding the two surfaces in contact under pressure of  $ \sim10^5 Pa $. Care is taken to ensure that no visible air pockets are trapped at the interface. In the case of soft top-hard bottom plate geometry, a parallel plate toolbar $(\rm PP50)$ of the rheometer is attached to the top surface of the PMMA disc.}
	\label{fig:expt_setup}
\end{figure}

\newpage

\begin{figure}[h]
	\centering
	\includegraphics[width=0.75\linewidth]{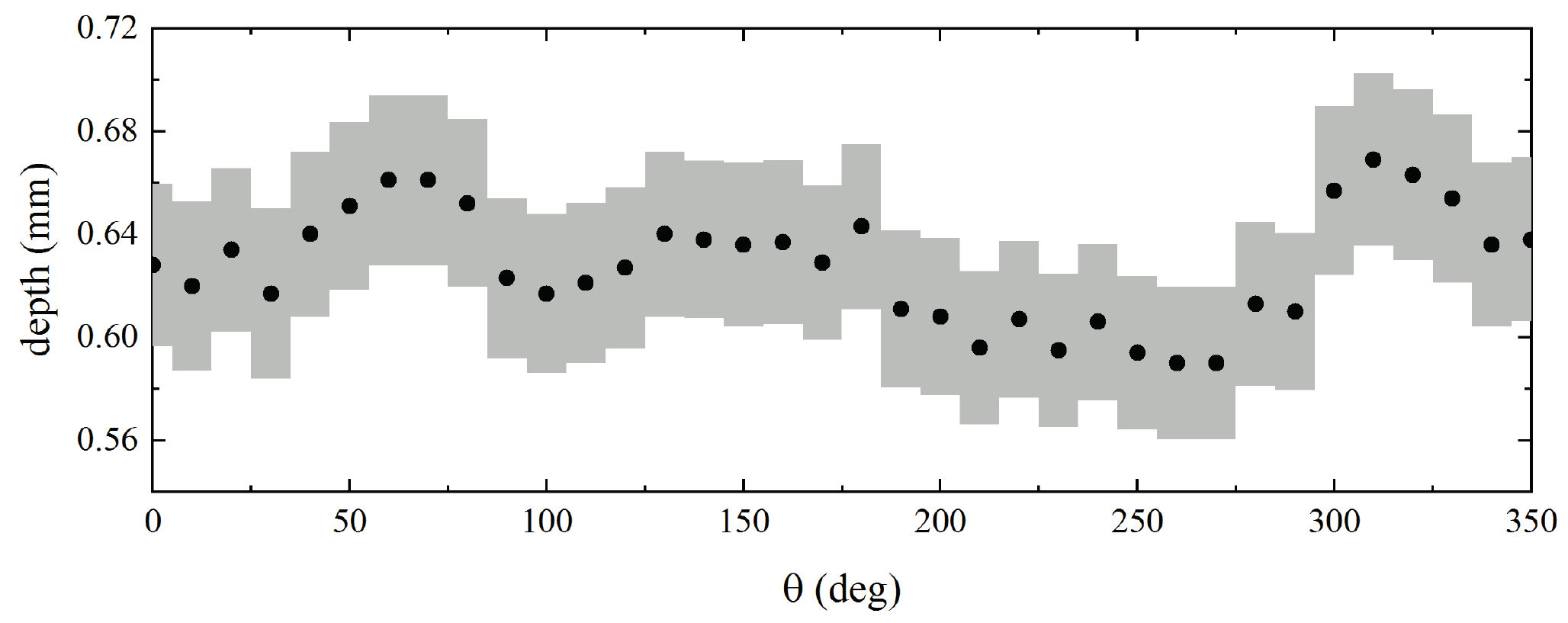}
	\caption{\textbf{Depth Profile of the Groove}: Due to unintended machining errors, the depth of the groove has an angular variation. This is shown in the above figure for the circular track of radius $47$ $\rm mm$. The average depth of the groove is $\sim 0.63$ $\rm mm$. }
	\label{fig:height profile}
\end{figure}

\newpage
\begin{figure}[h]
	\centering
	\includegraphics[width=0.75\linewidth]{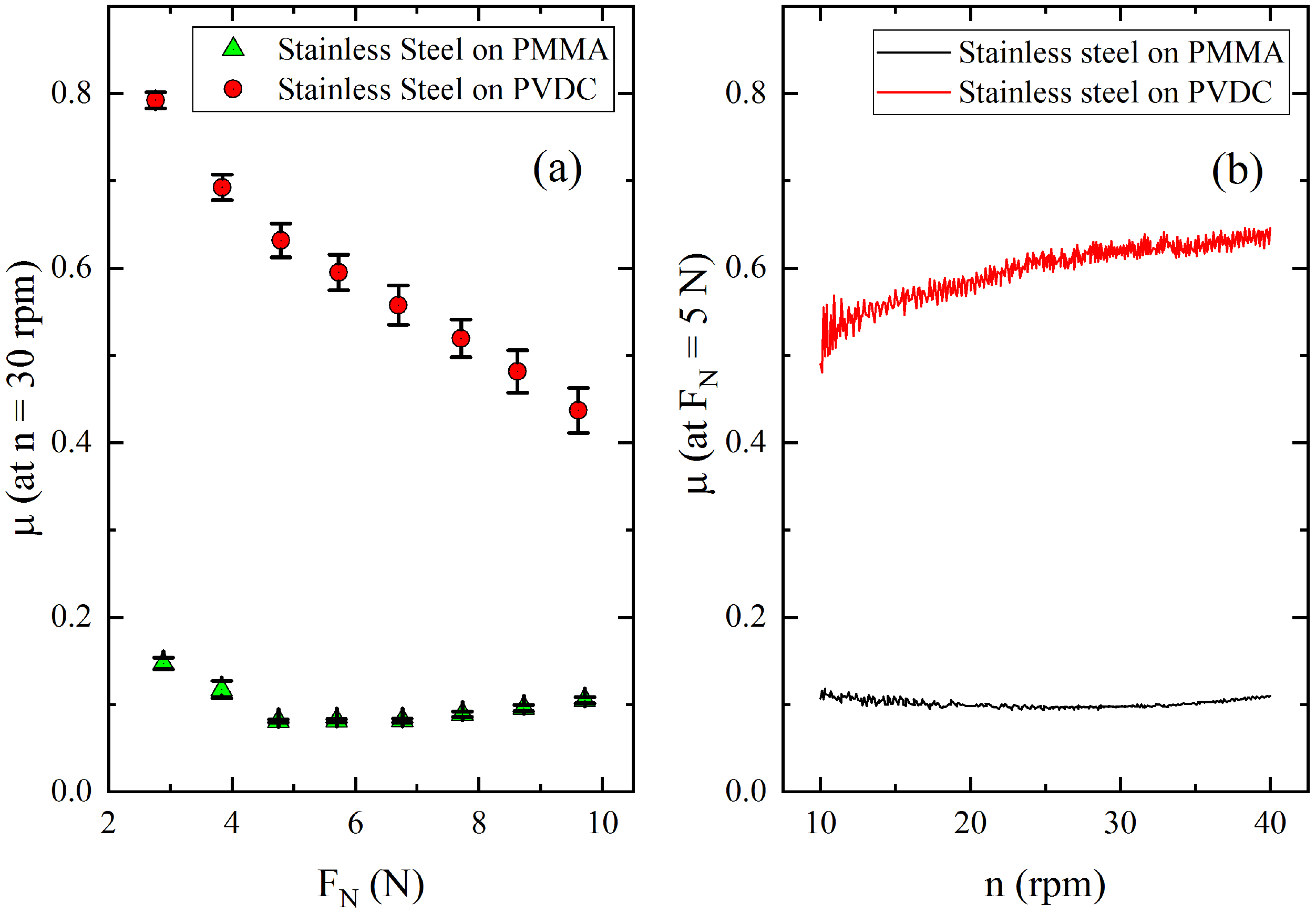}
	\caption{\textbf{Characterization of Sliding Friction}: The coefficients of sliding friction of the stainless steel ball on PMMA, PVDC, and stainless steel are measured using rheometer Physica MCR-301 in ball-on-three-plates geometry \cite{tribology_rheomter}. A measuring ball made of stainless steel is fixed to the measuring system shaft which is then pressed onto the three plates made of PMMA/PVDC sheet/stainless steel with a defined normal force $F_N$. From this normal force, normal load $F_L$ is calculated as $F_L=\frac{F_N}{\cos\alpha}$, where $\alpha$ is the angle made by the plates with the horizontal. The measuring ball is rotated at a constant angular speed of $n$ $\rm rpm$ which slides over the three plates. To maintain the set angular speed, a certain torque $\tau$ is required which is measured by the rheometer. The frictional force is calculated as $F_R=\frac{\tau}{r \sin\alpha}$, where $r=6.35$ $\rm mm$ is the radius of the measuring ball. The coefficient of friction is then $\mu=\frac{F_R}{F_L}=\frac{\tau}{r F_N \tan\alpha}$.  The panel (a) shows the variation of  $\mu$ as  a function of normal force for stainless steel on PMMAand on PVDC at a constant angular speed of $n=30$ $\rm rpm$.  The panel (b) shows the variation of the  $\mu$ as a function of angular speed for stainless steel on PMMAand on PVDC at a constant normal force of $F_N=5$ $\rm N$. The coefficient of sliding friction of stainless steel with itself is $\sim 0.57$. 
		}
	\label{fig:friction}
\end{figure}

 \newpage

\begin{figure}[h]
	
	\includegraphics[width=.6\linewidth]{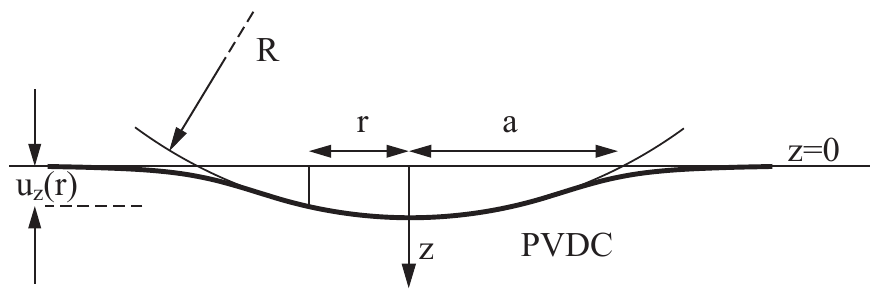}
	\caption{\leavevmode\\\begin{minipage}{\linewidth}
			\medskip
   		\textbf{Elastic Deformation of the soft plate}:
			The figure  shows a schematic representation of the side view of the contact zone when a steel ball is pressed onto the soft PVDC sheet in hard top-soft bottom plate geometry. A similar thing happens in the opposite geometry. The deformation of the soft plate can be modeled using Hertz contact theory \cite{johnson_1985,popov2010contact}. When the steel ball is pressed onto the PVDC sheet, pressure is exerted on a circle of radius $a$ as shown in the figure. Positive $z$ direction points into the PVDC sheet. The un-deformed surface of PVDC sheet is at $z=0$ plane and $R$ is the radius of the ball. We assume the pressure distribution at a distance $r$ from the center of contact circle is of Hertzian type, i.e., $p(r)=p_0\left(1-\frac{r^2}{a^2}\right)^{\frac{1}{2}}$, where $p_0=\left(\frac{6FE^2}{\left(1-\nu^2 \right)^2\pi^3R^2}\right)^{\frac{1}{3}}$ and $a=\left(\frac{3FR\left(1-\nu^2 \right)}{4E}\right)^{\frac{1}{3}}$. $E$ and $\nu$ are  Young's modulus and Poisson's ratio of PVDC respectively and $F$ is the applied normal force. The normal displacement at $r$ turns out to be \cite{johnson_1985}
		\begin{eqnarray*}
			\bar{u}_z(r) &=& \frac{\pi p_0(1-\nu^2)}{4aE}(2a^2-r^2), r\leq a \\
			&=& \frac{ p_0(1-\nu^2)}{2aE}\left[(2a^2-r^2)\sin^{-1}\left(\frac{a}{r}\right)+ra\left(1-\frac{a^2}{r^2}\right)^{\frac{1}{2}}\right] , r \geq a
		\end{eqnarray*} 
				The radial component of the stress in the surface $z=0$ is \cite{johnson_1985}
		\begin{eqnarray*}
			\bar{\sigma}_r(r) &=& \frac{p_0(1-2\nu)}{3}\frac{a^2}{r^2}\left[1-\left(1-\frac{r^2}{a^2}\right)^\frac{3}{2}-\left(1-\frac{r^2}{a^2}\right)^\frac{1}{2} \right] , r\leq a\\
			&=& \frac{p_0(1-2\nu)a^2}{3r^2},  r\geq a
		\end{eqnarray*}
				Therefore, outside the circle of contact, radial stress is tensile (i.e. pointing outward from the center of the contact circle) which causes repulsion between two nearby balls and prevents them from touching each other.
  \end{minipage}
		 }
	\label{fig:contact circle}
\end{figure}
\newpage

\begin{figure}[h]
	\centering
	\includegraphics[width=0.75\linewidth]{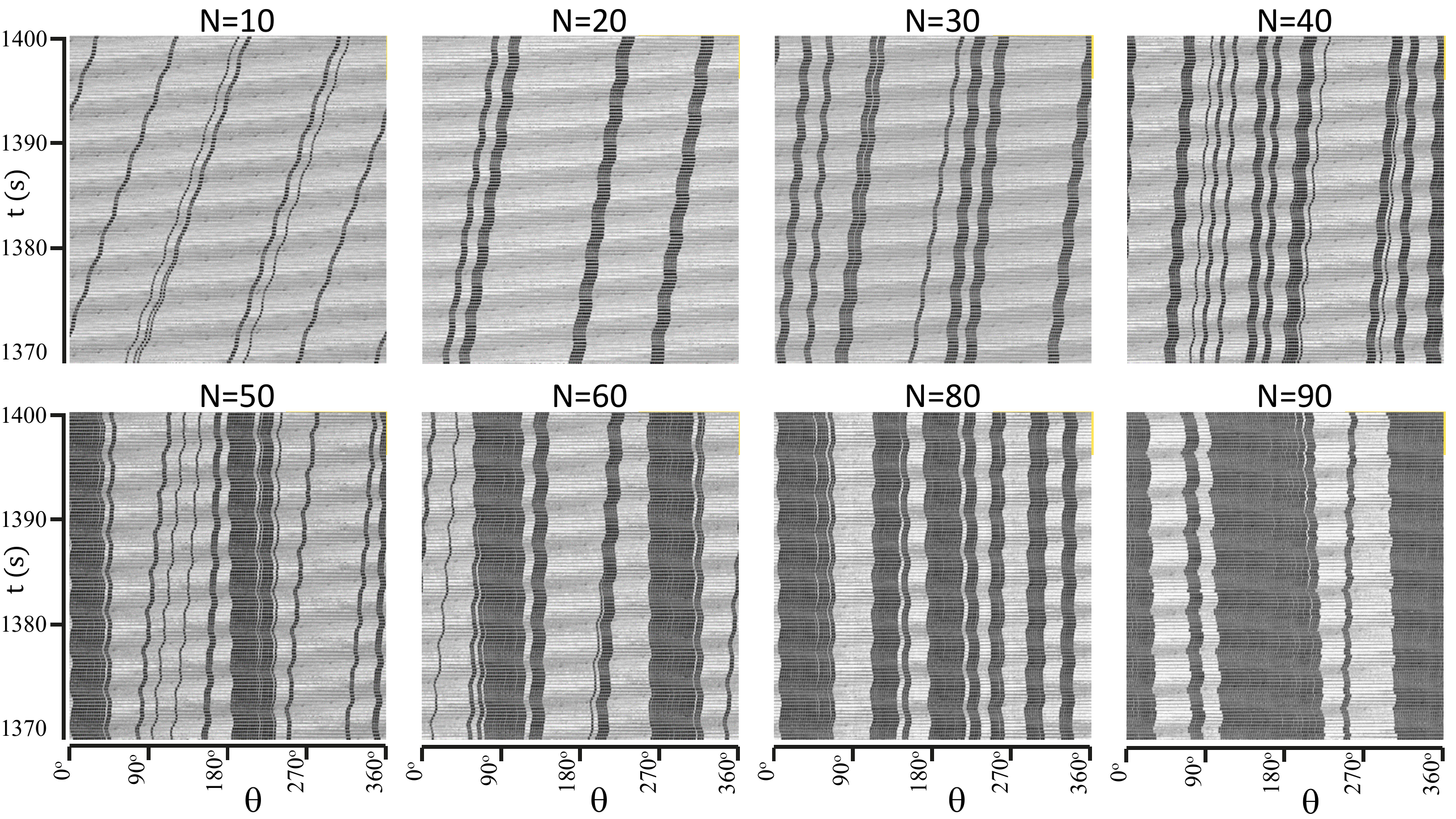}
	\caption{Montage of images for some representative values of $N$ in case of soft top-hard bottom late geometry for  values of $N$ other than those shown in the main manuscript. 
	}
	\label{fig:images}
\end{figure}

\newpage

\begin{figure}[h]
	\centering
	\includegraphics[width=0.75\linewidth]{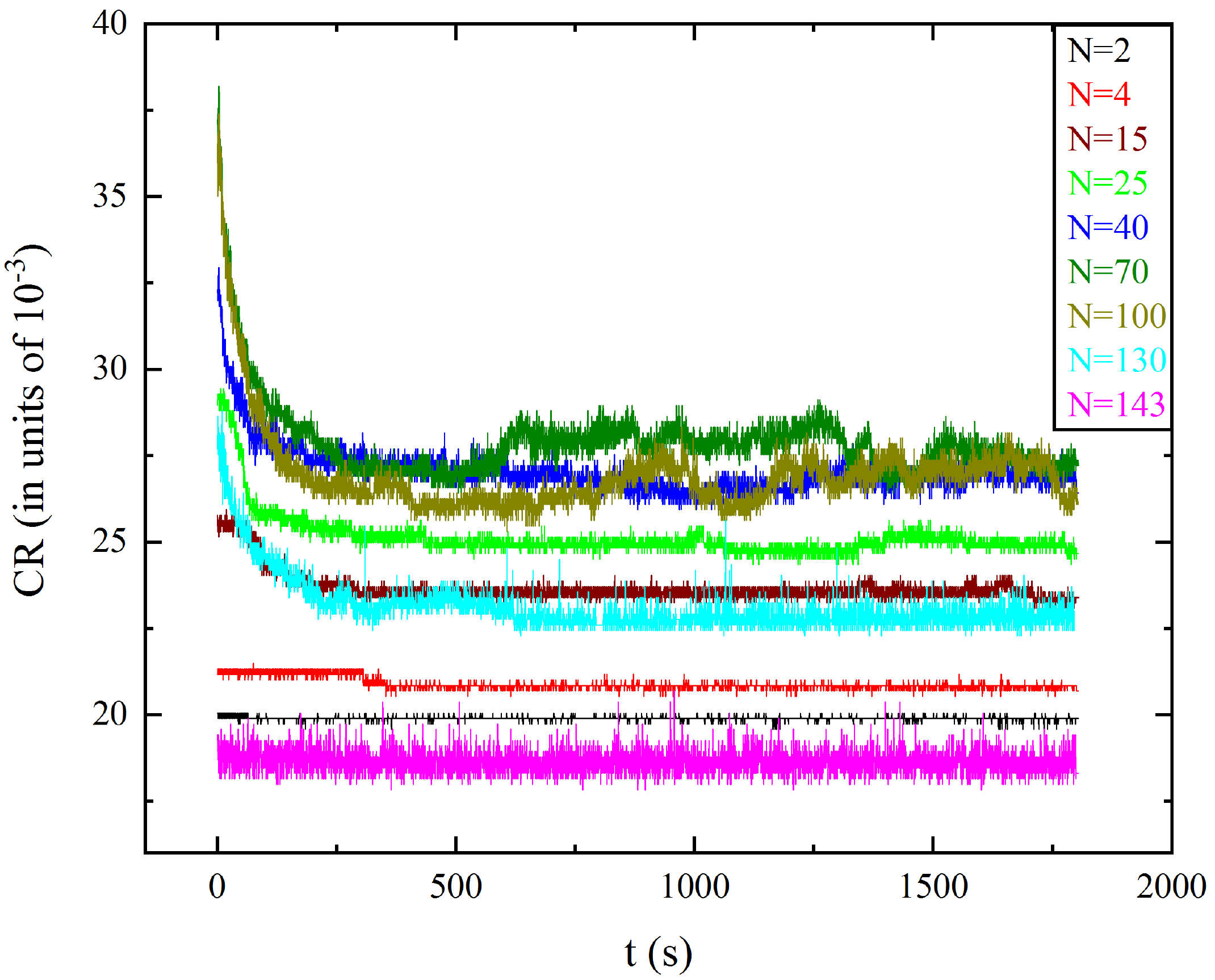}
	\caption{\textbf{LZW compression}: The variation of the compression ratio $CR$ as a function of time in the case of soft top-hard bottom plate geometry with track radius $R=47$ $\rm mm$ is shown in the figure for some representative values of $N$. For both very small and large $N$ (i.e. $N=2$ and $N=143$), the binary string corresponding to the spatial arrangement of the balls can be highly compressed due to the presence of a large number of contiguous '$0$'s and '$1$'s. Thus the $CR$ remains small and does not evolve in time. For other values of $N$, the $CR$ initially decreases and then varies about a steady-state value. The decrease in $ CR $ is related to the clustering of the balls and its long-time variations are related to the dynamic nature of the clusters.  By repeating the experiments for a given $\rho_N$  we observe that different initial configuration of the balls on the ring gives similar steady-state  $ CR$'s, i.e., the system does not show long term memory.  However, the time required to reach the steady-state shows a large spread as it depends strongly on the initial configuration.
}
	\label{fig:CR}
\end{figure}

\begin{figure}[t]
	\centering
	\includegraphics[width=.75\linewidth]{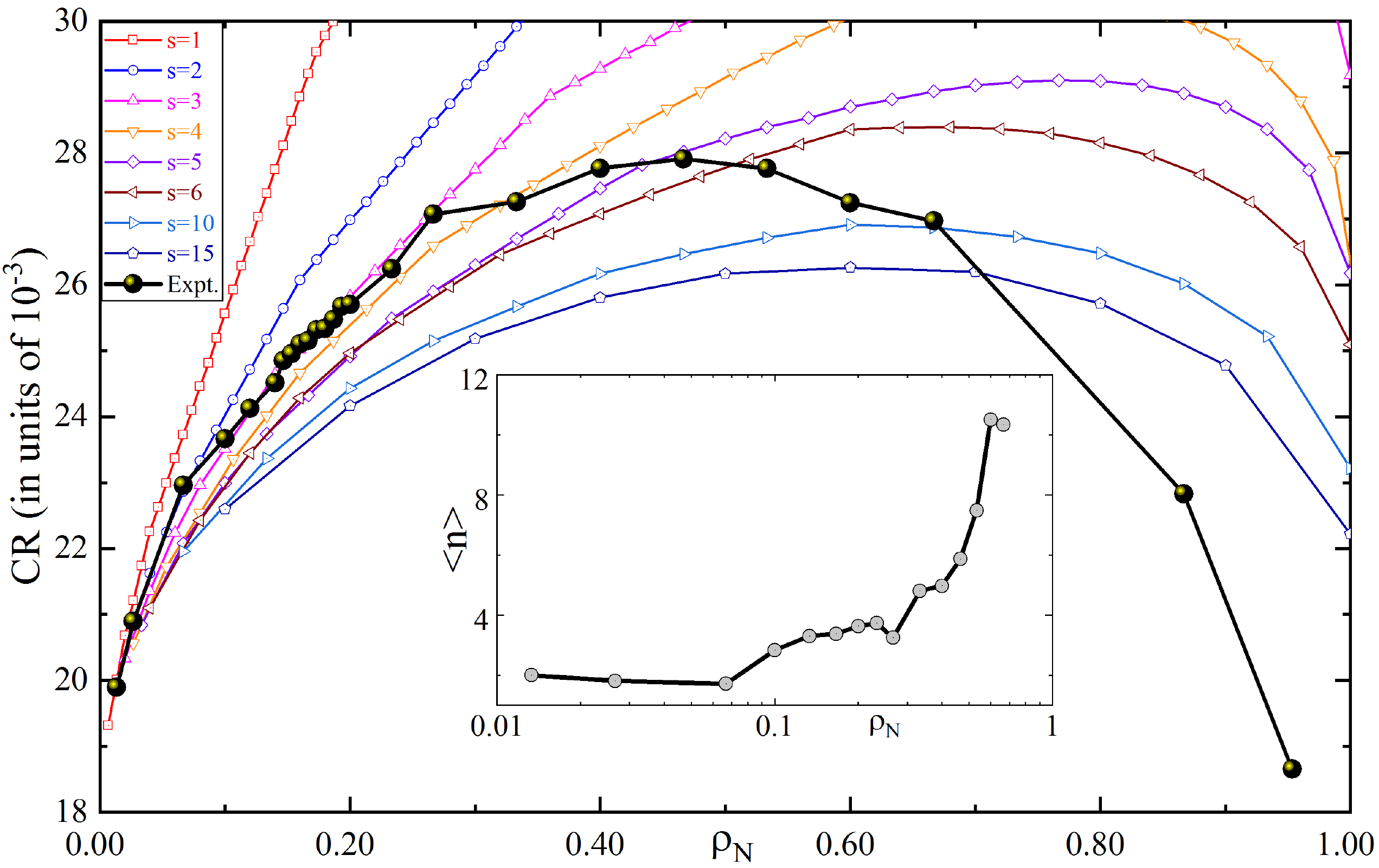}
	\caption{The steady-state part of $CR$ as a function of $\rho_N$ is shown in the scatter plots in the case of soft-top and hard-bottom plate geometry with $R=30$ $\rm mm$ and applied $F_N=30$ $\rm N$. The colored curves with open symbols are the simulated $CR$'s for different values of $s$. The mean value of $CR$ obtained from experiments are shown as black dots. Inset shows $\langle n \rangle$ as a function of $\rho_N$ determined from direct image analysis for soft-top and hard-bottom plate geometry.
	}
	\label{fig:fig5}
\end{figure}

\newpage
From the images in the soft-top and hard-bottom plate geometry, we obtain the average cluster size $\langle n \rangle$ as a function of $\rho_N$ (inset of Fig. \ref{fig:fig5}). This structural correlation can also be measured in terms of the extent by which the data can be compressed in a lossless manner. The usefulness of studying lossless data compression as a tool to find correlations in physical situations has been shown recently \cite{PhysRevX.9.011031}. If a physical system is more ordered, then the data string corresponding to the configuration of the system can be compressed more and vice versa. Here we compare the correlation length ($\langle n \rangle$) obtained from directly analyzing the images to that obtained from the lossless compression of data representing the images. For this purpose, the images are converted into binary sequences of length $ L $. Here $ L $ is equal to the width of the opened up image in pixels. In this binary sequence, each ball is replaced by a `block' of $ w (= L/N_T )$ $1$'s and the interstitial space is filled by $0$'s.  Fig. 3 (a) in the main manuscript is built from a stack of such images where $ L=6284 $ and $ w=41 $. These binary strings are then compressed using the lossless Lempel-Ziv-Welch (LZW) data compression algorithm and the compression ratio, $ CR $ is measured as the ratio of the length of the compressed sequence to that of the uncompressed sequence \cite{PhysRevX.9.011031}.

The black dots in Fig. \ref{fig:fig5} correspond to the long-time average base value of $CR$ of the experimentally obtained images for respective $\rho_N$'s. The curves with open symbols are generated as follows. For a given $\rho_N$, $1000$ random binary sequences are generated such that there are always $ s\times w $ number of contiguous $1$'s and the rest are $0$'s, where $s$ is a positive integer. We compute the average $CR$ of these $1000$ sequences. Thus for different values of $s$, different curves are generated as a function of $\rho_N$. Each curve then corresponds to the $CR$ of the data string corresponding to the system where all the clusters have the same size $ s$. The curve corresponding to the variation of the CR of the experimentally obtained images with $ \rho_N $ cross different branches of $ s $ marked by lines joining open symbols. The average cluster size can be extracted by seeing which branch of $s$, the experimental data lies on.
For $\rho_N \leq 0.7$, compression analysis matches with the average cluster size shown in the inset. However, for $\rho_N \geq 0.7$, compression data falls slightly short of that.

\newpage
\section*{LZW Compression Algorithm}

Lempel-Ziv-Welch (LZW) algorithm is  a commonly used lossless data compression algorithm  \cite{welch1984technique,sayood2002lossless,salomon2004data,lzw}. It starts out with a dictionary of $256$ characters (in the case of $8$ bits) and uses those as the standard character set. It then reads data $8$ bits at a time and encodes the data as the number that represents its index in the dictionary. Whenever it comes across a new substring, it adds it to the dictionary and whenever it comes across a substring it has already seen, it just reads in a new character and concatenates it with the current string to get a new substring. The next time LZW revisits a substring, it is encoded using a single number. Usually, a maximum number of entries (say, $4096$) is defined for the dictionary. Thus, the codes which are taking place of the substrings are 12 bits long ($2^{12} = 4096$). Since many frequently occurring substrings are replaced by a single code, compression is achieved. Below is the pseudocode - %\cite{lzw}-

\begin{algorithmic}
\State string $s$
\State char $ch$
\State $s =$ empty string
\While {there is still data to be read} 
\If {dictionary contains $s+ch$}
    \State $s\gets s+ch$
\Else
	\State encode $s$ to output file
    \State add $s+ch$ to dictionary
    \State $s\gets ch$
\EndIf
\EndWhile
\State encode $s$ to output file
\end{algorithmic}